\documentclass{cjaa}                   

\usepackage{graphicx}                   
\input{epsf.sty}                        
\input{psfig.sty}                       

\begin{document}

   \title{X-Ray Flares of Gamma-Ray Bursts:
   Quakes of Solid Quark Stars?
$^*$
}

   \volnopage{Vol.0 (200x) No.0, 000--000}      
   \setcounter{page}{1}          

   \author{Renxin Xu
      \inst{1}
  \and Enwei Liang     \inst{2}
      }
   \offprints{Renxin Xu}                   
   \institute{Astronomy Department, School of Physics, Peking University, Beijing 100871, China\\
%
        \and
            Department of Physics, Guangxi University, Nanning 530004, China \\
             \email{r.x.xu@pku.edu.cn, lew@gxu.edu.cn; {\rm Contributed by Renxin Xu.}}
          }
   \date{Received~~2008 month day; accepted~~2008~~month day}

   \abstract{
We propose a star-quake model to understand X-ray flares of both
long and short Gamma-ray bursts (GRBs) in a solid quark star regime.
Two kinds of central engines for GRBs are available if pulsar-like
stars are actually (solid) quark stars, i.e., the SNE-type GRBs and
the SGR-type GRBs. It is found that a quark star could be solidified
about $10^3$ to $10^6$ s later after its birth if the critical
temperature of phase transition is a few MeV, and then a new source
of free energy (i.e., elastic and gravitational ones, rather than
rotational or magnetic energy) could be possible to power GRB X-ray
flares.
   }

   \authorrunning{Xu \& Liang}            
   \titlerunning{GRB X-Ray Flares: Quakes of Solid Quark Stars?}  

   \maketitle

%
%
\section{Introduction}           

\label{sect:intro} {\em Swift},  a multi-wavelength gamma-ray burst
(GRB) mission (Gehrels et al. 2004), has led to great progress in
understanding the nature of the GRB phenomenon (see recent reviews
by M\'{e}sz\'{a}ros 2006; Zhang 2007). With its promptly slewing
capacity and high sensitivity, it catches the early afterglows and
the extended prompt emission in details for the first time. This not
only provides an opportunity to examine the conventional models
established in the pre-Swift era (Willingale et al. 2007; Liang et
al. 2007, 2008; Panaitescu 2007; Zhang et al. 2007), but also
facilitates studies of the transition between the prompt emission
and the afterglow (Zhang et al. 2007,2008; Butler \& Kocevisky
2007), and even gives insight into the properties of both the
progenitors and the GRB central engines (e.g., Liang et al. 2007;
Kumar et al. 2008).

The GRB survey with CGRO (Compton Gamma-ray Observatory)/BATSE
identified two types of GRBs, long-soft and short-hard GRBs,
separated with burst duration of $\sim 2$ seconds (Kouveliotou et
al. 1993). On one hand, with firmed detections of GRB-supernovea
connections for four nearby cases (Galama et al. 1998; Bloom et al.
1999; Hjorth et al. 2003; Thomsen et al. 2004; Campana et al. 2006),
it is now generally accepted that long GRBs are associated with
energetic core-collapse supernovae (Colgate 1974; Woosley 1993; for
recent reviews by Woosley \& Bloom 2006). Interestingly, Li (2006)
found that the peak spectral energy of GRBs is correlated with the
peak bolometric luminosity of the underlying supernovae (SNe), based
on the four pair GRB-SNe connections. The X-ray transient 080109
associated with a normal core-collapse SN 2008D (Soderberg et al.
2008) also complies with this relation (Li 2008). Signatures of long
GRB-SNe connection may be also derived from a red bump in late
optical afterglow lightcurves (Bloom et al. 1999; Zeh, Klose, \&
Hartmann 2004) and a long time lag between the GRB precursor and the
main burst observed in some GRBs (Wang \& M{\'e}sz{\'a}ros 2007). On
the other hand, short GRBs coincide with the early-type stellar
population with no or little current star formation (Gehrels et al.
2005; Berger et al. 2005; Barthelmy et al. 2005; Hjorth et al. 2005;
Villasenor et al. 2005; Fox et al. 2005; see recent review by Nakar
2007), favoring mergers of compact object binaries as the
progenitors of the short GRBs (Goodman 1986; Eichler et al. 1989;
Paczynski 1991; M\'{e}sz\'{a}ros \& Rees 1992; Narayan, Paczynski,
\& Piran 1992).

Although the progenitors of the long and short GRBs are different,
the models for their central engines are similar, and essentially
all can be simply classed as a rotating compact object that drives
an ultra-relativistic outflow to produce both the prompt gamma-rays
and afterglows in lower energy bands. These models are highly
constrained by the observations of the prompt gamma-rays and
multi-wavelength afterglows. It is well believed that the prompt
gamma-rays are produced by the internal shocks, and the burst
duration is a measure of the central engine active timescale. In the
merger models of compact object binaries for the short GRBs, the
duration of the bursts is expected to be less than 1 second
(Narayan, Piran, \& Kumar 2001). This is challenged by the
observations with  {\em Swift}. One of the remarkable advances made
by {\em Swift} is the discovery of erratic X-ray flares for both
long and short GRBs, happening at very early time or hours even one
day after the GRB trigger in the light curves observed with the
X-ray telescope (XRT) (Burrows et al. 2005; Falcone et al. 2006;
Chincarini et al. 2007).  The X-ray flares are a superimposed
component of the underlying afterglows, with a feature of rapid rise
and fall times ($\delta t<< t_{peak}$). Multiple flares are observed
in some bursts. They are similar to the pulse of the prompt
gamma-rays, but the fluence of the flares decrease with time and the
durations of the flares at later time become broader than early
flares. These properties generally favor the idea that most of them
are of internal origin, having nothing to do with external-shock
related events (Burrows et al. 2005; Fan \& Wei 2005; Zhang et al.
2006; Liang et al. 2006). This indicates that the central engine
should live much longer than the burst duration or it was
re-restarted by an un-recovered mechanism. Several models have been
proposed, such as magnetic explosions of a millisecond pulsar from
NS-NS merger (Dai et al. 2006), fragmentation or gravitational
instabilities in the massive star envelopes (King et al. 2005) or in
the accretion disk (Perna et al. 2006), or magnetic barrier around
the accretor (Proga \& Zhang 2006).

The facts that X-ray flares are observed for both long and short
GRBs motivate us to speculate that the central engines of the two
kinds of GRBs are physically similar. We know that an accretor-disk
system is hard to sustain a long lived engine for the short GRBs
(Narayan, Piran, \& Kumar 2001), except a fraction of materials is
lunched to a large orbit. However, the fall-back of the materials
cannot produce the observed erratic flares (Rosswog 2007). We
therefore propose alternatively that the mechanism should be
harbored in the central star (i.e., the engine).

Actually, the essential difficulty of reproducing two kinds of
astronomical bursts are challenging today's astrophysicists to find
realistic explosive mechanisms. Besides the puzzling center engines
of GRBs, it is still a long-standing problem to simulate supernovae
successfully in the neutrino-driven explosion model (e.g., Buras et
al. 2003). Nevertheless, it is shown now that both kinds of
explosions could be related to the physics of cold matter at
supra-nuclear density, which is unfortunately {\em not} well
understood because of the uncertainty of non-perturbative QCD
(quantum chromo-dynamics). We still do not know the nature of
pulsars with certainty even more than 40 years after the discovery,
which is also relevant to the physics of cold dense matter. Nuclear
matter (related to {\em neutron} stars) is one of the speculations,
but quark matter (related to {\em quark} stars) is an alternative
(e.g., Weber 2005). In this paper, we will speculate about the
physical reasons that make the mechanisms for both long-soft and
short-hard GRBs, which are related to the elemental strong
interaction, says, the physics of the cold dense mater at
supra-nuclear density. We suggest here that GRB X-ray flares could
be the results of star quakes of solid quark stars (Xu 2003; Owen
2005; see, e.g., Xu 2008 for a general review about quark stars).
Our idea for understanding GRB X-ray flares is presented in section
2, and the paper is summarized in section 3.

\section{An idea of understanding GRB X-ray flares}

Based on different manifestations of pulsar-like stars, a conjecture
of solid cold quark matter was addressed a few years ago (Xu 2003).
Consequently, a variety of observational features, which may
challenge us in the hadron star model, could be naturally understood
in the solid quark star model (Xu 2008), including the giant flares
of soft gamma-ray repeaters (Horvath 2005; Xu 2007).

What if pulsar-like stars are actually quark stars?
One of the direct and important consequences could be the low
baryon-loading energetic fireballs formed soon after quark stars,
which would finally result in both supernova and GRBs.
As addressed in Xu (2005), the {\em bare} quark surfaces could be
essential for successful explosions of both types of core and
accretion-induced collapses. The reason is that, because of the
strong binding of baryons, the photon luminosity of a quark surface
is not limited by the Eddington limit, and it is thus possible that
the prompt reverse shock could be revived by photons, rather than by
neutrinos.
This point was then noted too by Paczy\'{n}ski and Haensel (2005)
who proposed that classical long-duration gamma-ray bursts could be
from the formation of quark stars several minutes after the initial
core collapse, emphasizing the surface as a {\em membrane} allowing
only ultrarelativistic non-baryonic matter to escape.
Actually, a 1-dimensional (i.e., spherically symmetric) calculation
by Chen, Yu \& Xu (2007) showed that the lepton-dominated fireball
supported by a bare quark surface do play a significant role in the
explosion dynamics under a photon-driven scenario.
Recently, the QCD phase transition for quark matter during the
post-bounce evolution of core collapse supernovae was numerically
investigated by Sagert et al. (2008), and they found that the phase
transition produces a second shock wave that triggers a delayed
supernova explosion.
However, what if the expanding of a fireball outside quark surface
is not spherically symmetric?
An asymmetric explosion may result both in a GRB-like fire jet and
in a kick on quark stars, and the statistical result of Cui et al.
(2007) indicated that the kick velocity of pulsars could be
consistent with an asymmetric explosion of GRBs.

Another consequence of quark star in a solid state is spontaneous
quake occurring when elastic energy develops to a critical value
there, if cold quark matter is actually in a solid state.
A nascent quark star could be in a fluid state since quarks are just
de-confined from hadrons, though it is still not sure whether the
quarks are clustered in a fluid state initially.
Anyway such a quark star should have to be solidified soon due
strong cooling through both neutrino and photon, and we may use a
toy model to estimate the timescale for a transition from fluid to
solid states.

For the sake of simplicity, we may approximate a quark star as a
star with homogenous density of $\rho=3\rho_0$ ($\rho_0\simeq
2\times 10^{14}$ g/cm$^{-3}$ is the nuclear density) and in a radius
of $R$, and its mass is then $M=4\pi R^3\rho/3 = 1.3 R_6^3M_\odot$,
where $R=R_6\times 10^6$ cm.
The total quark number is $4.5\times 10^{57}$, and the total number
of quark clusters could be $N_{\rm qc}=4.5 \times 10^{56}R_6^3$ if
the average quark number in clusters is order of 10.
The total thermal energy of a quark star is then
\begin{equation}
Q={3\over 2}k T N_{\rm qc}=1.1\times 10^{51}T_1R_6^3\; {\rm ergs},
\end{equation}
where the stellar temperature $T(t)=T_1$ MeV is assumed to be
constant at certain age $t$.

In high temperature, neutrino emissivity via pair annihilation
($\gamma + \gamma \leftrightarrow e^\pm \rightarrow \bar{\nu} +
\nu$) dominates. We just consider this mechanism for neutrino
cooling since the $\nu$-emissivity of clustered quark matter is
hitherto unknown. The emissivity was presented by Itoh et al. (1989)
who used the Weinberg-Salam theory in their calculation, which is.
\begin{equation}
\epsilon_{\rm
pair}=1.089[1+0.104q(\lambda)]g(\lambda)e^{-2/\lambda}f(\lambda)\;
{\rm erg\; s^{-1}\; cm^{-3}},
\end{equation}
where $\lambda=T/(5.9302\times 10^9$ K), $\xi=[\rho/\mu_e/(10^9~{\rm
g/cm^3})]^{1/3}\lambda^{-1}$, and,
$$
q(\lambda) =
(10.7480\lambda^2+0.3967\lambda^{0.5}+1.0050)^{-1.0}[1+(\rho/\mu_e)(7.692\times
10^7\lambda^3+9.715\times 10^6\lambda^{0.5})^{-1.0}]^{-0.3},$$
$$g(\lambda)=1-13.04\lambda^2+133.5\lambda^4+1534\lambda^6+918.6\lambda^8,$$
$$f(\lambda)={(a_0+a_1\xi+a_2\xi^2)e^{-c\xi}\over
\xi^3+b_1\lambda^{-1}+b_2\lambda^{-2}+b_3\lambda^{-3}},
$$
where $a_0 = 6.002\times 10^{19}$, $a_1 = 2.084\times 10^{20}$, $a_2
= 1.872\times 10^{21}$; $b_1 = 0.9383$,  $b_2 = -0.4141$,  $b_3 =
0.05829$, $c = 5.5924$ for $T<10^{10}$ K; $b_1 = 1.2383$,  $b_2 =
-0.8141$,  $b_3 = 0.0$, $c = 4.9924$ for $T\geq 10^{10}$ K.
According to the calculation by Zhu \& Xu (2004) in the bag model,
the ratio of number density of electron to that of quark is
$<10^{-4}$. We thus choose the electron mean molecular weight
$\mu_e=10^5$ for strange quark matter in following calculation.

Assuming the optical depth of neutrinos in proto-quark stars is less
than 1, we then have the energy loss rate due to neutrino emission,
\begin{equation}
{\dot Q}_\nu\simeq {4\over 3}\pi R^3\epsilon_{\rm pair}.
\end{equation}
Another important cooling mechanism for {\em bare} quark stars is
thermal photon emission from quark surface,
\begin{equation}
{\dot Q}_\gamma\simeq 4\pi R^2 \sigma T^4,
\end{equation}
and the total cooling rate is then,
\begin{equation}
{\dot Q}={3\over 2}kN_{\rm qc}{{\rm d}T\over {\rm d}t}=-{\dot
Q}_\nu-{\dot Q}_\gamma.%
\label{cooling}
\end{equation}
A comparison of ${\dot Q}_\nu$ and ${\dot Q}_\gamma$ is shown in
Fig.1.
\begin{figure}
  \centering
    \includegraphics[width=8cm]{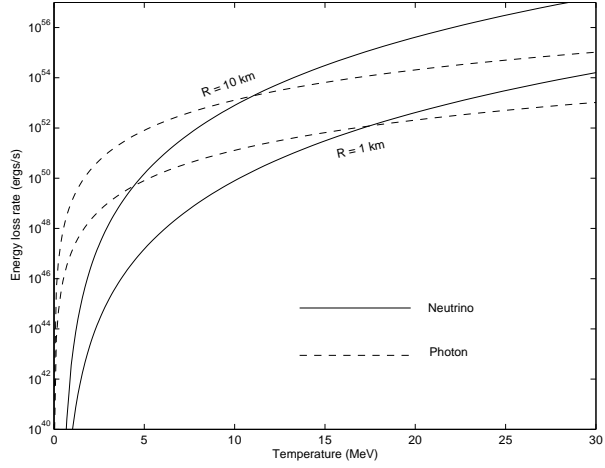}
    \caption{%
Cooling rates due to neutrino and photon emissivities for quark
stars with radii of 10 km (solid lines) and 1 km (dashed lines). It
is evident that photon emissivity dominates except at the very
beginning of stars with high temperature, and the critical
temperature is higher for quark stars with lower masses.
\label{f1}}
\end{figure}
It is evident that the neutrino loss dominated at high temperatures,
while the photon emissivity dominates at low temperatures. The
critical temperature $T_{\rm crit}$, at which the energy loss rates
of neutrinos and photons are equal, depends on stellar radius (or
mass). Low mass quark stars have higher $T_{\rm crit}$.

According to Eq.(\ref{cooling}), we can also calculate the cooling
curves of quark stars in the toy model, which is shown in Fig.2.
\begin{figure}
  \centering
    \includegraphics[width=8cm]{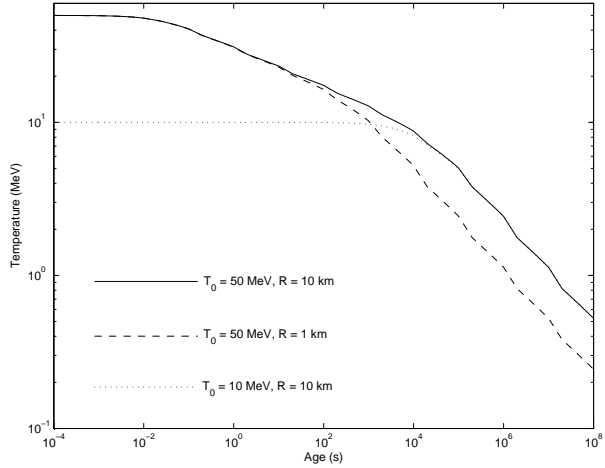}
    \caption{%
Cooling curves based on Eq.(\ref{cooling}) for initial temperature
$T_0=50, 10$ MeV and stellar radii of 10, 1 km. For a star with
$T_0=10$ MeV and $R=1$ km, its cooling curve fits the dotted line
before age $t\sim 10^2$ s, but fits the dashed line after $t\sim
10^3$ s.
\label{f1}}
\end{figure}
It was suggested that quark-clusters in cold quark matter could be
localized at lattices, breaking then the translational invariance,
to form solid quark matter with rigidity (Xu 2003), but we are not
sure about the critical temperature, $T_c$, at which the
solidification phase transition happens because of lacking reliable
way to calculate with the inclusion of non-perturbative QCD effects.
Nevertheless, a quark star could be solidified about $10^3$ to
$10^6$ s later after its birth if the critical temperature is $T_c
\sim 1$ to 10 Mev, as is illustrated in Fig.2.
That $T_c$ is order of a few MeV could be reasonable and not
surprising since the interacting strength of nuclei, where the
non-perturbative QCD effects dominate, is also of this energy scale.

What if quark stars are in a solid state?
Star-quake is a natural consequence when strains accumulate to a
critical value.
Two types of stress force could develop inside solid stars (Peng \&
Xu 2008): the bulk-variable and bulk-invariable forces. Both these
forces could result in gravitational and elastic energy releases (to
be in a same order), with an order of
\begin{equation}
E\simeq {GM^2\over R} \sim 10^{53}{\Delta R\over R}\; {\rm ergs}
\end{equation}
for $M\sim M_\odot$, where $\Delta R$ is the radius-change during
quakes. An energy release of $\sim 10^{50}$ ergs could be possible
if stellar radius changes suddenly $\sim 10$ m during a quake of a
solid quark star with radius of $\sim 10$ km.
The rise time of a burst could be $>{\hat t} \sim {\hat R}/c\sim 1$
ms, where ${\hat R}$ is the scale of a quake-induced fireball in a
magnetosphere and could be a few tens of stellar radius. The rise
time would be $\gg{\hat t}$ if energy is ejected into fireball by a
series of small quakes. The duration would actually depend on detail
radiative process in magnetosphere.

In the regime of quark star, there could be two kinds of mechanisms
for $\gamma$-ray bursts.
The first one, we call as ``SNE-type'', is relevant to the birth of
quark stars and supernovae.
A very clean (i.e., lepton-dominated) fireball forms soon above
quark surface. In addition, more energy would be ejected into the
fireball when star-quakes occur.
The discovery of erratic X-ray flares in long-soft GRBs, happening
at very early time or hours even one day after the GRB trigger, is
consistent with this picture.

The second one, we call as ``SGR-type'', is relevant to the later
evolution of quark stars, especially in an accretion phase.
An accretion-induced star-quake (AIQ) model was suggested to
understand the huge energy bursts of soft Gamma-ray repeaters
(SGRs), based on several calculations of the static, spherically
symmetric, and interior solution (Xu et al. 2006, Xu 2007, Lai \& Xu
2008).
It is found that the energy released during star-quakes could be as
high as $\sim 10^{47}$ ergs if the tangential pressure is $\sim
10^{-6}$ higher than the radial one.
A big star-quake could power energetic relativistic outflow to
produce the observed prompt emission of short-hard GRBs, and this
quake may trigger a few smaller and stochastic quakes which result
in following X-ray flares observed.

\section{Conclusions}

An idea to understand the X-ray flares of both long and soft
Gamma-ray bursts is proposed in the quark star regime. We suggest an
SNE-type GRB scenario when quark stars are born, and an SGR-type GRB
scenario if giant quakes occur in solid quark stars during their
latter accretion phases.
However, stochastic quakes after initial GRBs could be responsible
to the X-ray flares of both types of GRBs.

According to a toy model of cooling quark stars, we find that a
quark star could be solidified about $10^3$ to $10^6$ s later after
its birth if the critical transition temperature is $\sim 1$ to 10
MeV. This means that star-quakes could occur at a time of hours
(even one day) after the GRB trigger, and the star-quake induced
energy ejection would then results in the observed X-ray flares of
SNE-type GRBs.

\begin{acknowledgements}
We acknowledge valuable discussion with Dr. Lixin Li (KIAA) and Dr.
Bing Zhang (UNLV). This work is supported by NSFC (10573002,
10778611, 10873002), the Key Grant Project of Chinese Ministry of
Education (305001), the National Basic Research Program (``973''
Program) of China (Grant 2009CB824800), the research foundation of
Guangxi University, and the LCWR (LHXZ200602).
\end{acknowledgements}

\label{lastpage}


\begin{thebibliography}{99}
\bibitem[Barthelmy et al.(2005)]{2005Natur.438..994B} Barthelmy, S.~D., et al.\ 2005, Nature, 438, 994
\bibitem[Berger et al.(2005)]{2005Natur.438..988B} Berger, E., et al.\ 2005, Nature, 438, 988
\bibitem[Bloom et al.(1999)]{1999Natur.401..453B} Bloom, J.~S., et al.\ 1999, Nature, 401, 453
\bibitem[Burrows et al. 2005]{2005Sci...309.1833B} Burrows, D.~N., et al.\ 2005, Science, 309, 1833
\bibitem[Butler \& Kocevski(2007)]{2007ApJ...668..400B} Butler, N.~R., \& Kocevski, D.\ 2007, ApJ, 668, 400
\bibitem{}Buras, R., Rampp, M., Janka, H.-Th., \& Kifonidis, K., 2003, Phys. Rev. Lett., 90, 241101
\bibitem[Campana et al.(2006)]{2006Natur.442.1008C} Campana, S., et al.\ 2006, Nature, 442, 1008
\bibitem{} Chen, A. B., Yu, T. H.,  Xu, R. X., 2007, ApJ, 668, L55
\bibitem{} Cui, X. H., Wang, H. G., Xu, R. X., Qiao, G. J., 2007, A\&A, 472, 1
\bibitem[Chincarini et al.(2007)]{2007ApJ...671.1903C} Chincarini, G., et al.\ 2007, ApJ, 671, 1903
\bibitem[Colgate(1974)]{1974ApJ...187..333C} Colgate, S.~A.\ 1974, ApJ, 187, 333
\bibitem[Dai et al.(2006)]{2006Sci...311.1127D} Dai, Z.~G., Wang, X.~Y., Wu, X.~F., \& Zhang, B.\ 2006, Science, 311, 1127
\bibitem[Eichler et al.(1989)]{1989Natur.340..126E} Eichler, D., Livio, M., Piran, T., \& Schramm, D.~N.\ 1989, Nature, 340, 126
\bibitem[Falcone et al.(2006)]{2006ApJ...641.1010F} Falcone, A.~D., et al.\ 2006, ApJ, 641, 1010
\bibitem[Fan \& Wei(2005)]{2005MNRAS.364L..42F} Fan, Y.~Z., \& Wei, D.~M.\ 2005, MNRAS, 364, L42
\bibitem[Fox et al.(2005)]{2005Natur.437..845F} Fox, D.~B., et al.\ 2005, Nature, 437, 845
\bibitem[Galama et al.(1998)]{1998Natur.395..670G} Galama, T.~J., et al.\ 1998, Nature, 395, 670
\bibitem[Gehrels et al. 2004]{2004ApJ...611.1005G} Gehrels, N., et al.\ 2004, ApJ, 611, 1005
\bibitem[Gehrels et al.(2005)]{2005Natur.437..851G} Gehrels, N., et al.\ 2005, Nature, 437, 851
\bibitem[Goodman(1986)]{1986ApJ...308L..47G} Goodman, J.\ 1986, ApJ, 308, L47
\bibitem[Hjorth et al.(2003)]{2003Natur.423..847H} Hjorth, J., et al.\ 2003, Nature, 423, 847
\bibitem[Hjorth et al.(2005)]{2005Natur.437..859H} Hjorth, J., et al.\ 2005, Nature, 437, 859
\bibitem{} Horvath, J. 2005, Mod. Phys. Lett., A20, 2799
\bibitem{} Itoh, N., Adachi, T., Nakagawa, M., Kohyama, Y., Munakata, H., 1989, ApJ, 339, 354
\bibitem[King et al.(2005)]{2005ApJ...630L.113K} King, A., O'Brien, P. T., Goad, M. R., Osborne, J., Olsson, E., \& Page, K.\ 2005, ApJ, 630, L113
\bibitem[Kouveliotou et al.(1993)]{1993ApJ...413L.101K} Kouveliotou, C., Meegan, C.~A., Fishman, G.~J., Bhat, N.~P., Briggs, M.~S., Koshut, T.~M., Paciesas, W.~S., \& Pendleton, G.~N.\ 1993, ApJ, 413, L101
\bibitem[Kumar et al.(2008)]{2008Sci...321..376K} Kumar, P., Narayan, R., \& Johnson, J.~L.\ 2008, Science, 321, 376
\bibitem{} Lai, X. Y., Xu, R. X., 2008, preprint (arXiv:0804.0983)
\bibitem[Li(2006)]{2006MNRAS.372.1357L} Li, L.-X.\ 2006, MNRAS, 372, 1357
\bibitem[Li(2008)]{2008MNRAS.388..603L} Li, L.-X.\ 2008, MNRAS, 388, 603
\bibitem[Liang et al.(2006)]{2006ApJ...646..351L} Liang, E.~W., et al.\ 2006, ApJ, 646, 351
\bibitem[Liang et al.(2007)]{2007ApJ...670..565L} Liang, E.-W., Zhang, B.-B., \& Zhang, B.\ 2007, ApJ, 670, 565
\bibitem[Liang et al.(2008)]{2008ApJ...675..528L} Liang, E.-W., Racusin, J.~L., Zhang, B., Zhang, B.-B., \& Burrows, D.~N.\ 2008, ApJ, 675, 528
\bibitem[M{\'e}sz{\'a}ros 2006]{2006RPPh...69.2259M} M{\'e}sz{\'a}ros, P.\ 2006, Reports of Progress in Physics, 69, 2259
\bibitem[Meszaros \& Rees(1992)]{1992ApJ...397..570M} M{\'e}sz{\'a}ros, P., \& Rees, M.~J.\ 1992, ApJ, 397, 570
\bibitem[Nakar(2007)]{2007PhR...442..166N} Nakar, E.\ 2007, \physrep, 442, 166
\bibitem[Narayan et al.(1992)]{1992ApJ...395L..83N} Narayan, R., Paczynski, B., \& Piran, T.\ 1992, ApJ, 395, L83
\bibitem[Narayan et al.(2001)]{2001ApJ...557..949N} Narayan, R., Piran, T., \& Kumer, P.\ 2001, ApJ, 557, 949
\bibitem{} Owen, B. J. 2005, Phys. Rev. Lett., 95, 211101
\bibitem[Paczynski & Haensel(2005)]{2005MNRAS.362L...4P} Paczy{\' n}ski, B., \& Haensel, P.\ 2005, MNRAS, 362, L4
\bibitem[Panaitescu(2007)]{2007MNRAS.380..374P} Panaitescu, A.\ 2007, MNRAS, 380, 374
\bibitem{} Peng, C.,  Xu, R. X., 2008, MNRAS, 384, 1034
\bibitem[Perna et al.(2006)]{2006ApJ...636L..29P} Perna, R., Armitage, P.~J., \& Zhang, B.\ 2006, ApJ, 636, L29
\bibitem[Proga \& Zhang(2006)]{2006MNRAS.370L..61P} Proga, D., \& Zhang, B.\ 2006, MNRAS, 370, L61
\bibitem[Rosswog(2007)]{2007MNRAS.376L..48R} Rosswog, S.\ 2007, MNRAS, 376, L48
\bibitem{} Sagert, I., Hempel, M., Pagliara, G., Schaffner-Bielich, J., Fischer, T., Mezzacappa, A., Thielemann, F.-K., \& Liebend{\" o}rfer, M.\ 2008, preprint (arXiv:0809.4225)
\bibitem[Soderberg et al.(2008)]{2008Natur.453..469S} Soderberg, A.~M., et al.\ 2008, Nature, 453, 469
\bibitem[Thomsen et al.(2004)]{2004AA...419L..21T} Thomsen, B., et al.\ 2004, A \& A, 419, L21
\bibitem[Villasenor et al.(2005)]{2005Natur.437..855V} Villasenor, J.~S., et al.\ 2005, Nature, 437, 855
\bibitem{} Weber, F., 2005, Prog. Part. Nucl. Phys., 54, 193
\bibitem[Wang \& M{\'e}sz{\'a}ros(2007)]{2007ApJ...670.1247W} Wang, X.-Y., \& M{\'e}sz{\'a}ros, P.\ 2007, ApJ, 670, 1247
\bibitem[Willingale et al.(2007)]{2007ApJ...662.1093W} Willingale, R., et al.\ 2007, ApJ, 662, 1093
\bibitem[Woosley \& Bloom(2006)]{2006ARA&A..44..507W} Woosley, S.~E., \& Bloom, J.~S.\ 2006, ARA\&A, 44, 507
\bibitem[Woosley(1993)]{1993ApJ...405..273W} Woosley, S.~E.\ 1993, ApJ, 405, 273
\bibitem{} Xu, R. X., Tao, D. J., Yang, Y., 2006, MNRAS, 373, L85
\bibitem{} Xu, R. X., 2003, ApJ, 596, L59
\bibitem{} Xu, R. X. 2005, MNRAS, 356, 359
\bibitem{} Xu, R. X., 2007, Adv. Spa. Res., 40, 1453
\bibitem{} Xu, R. X., 2008, Modern Phys. Lett., A23, 1629
\bibitem{} Zeh, A., Klose, S., \& Hartmann, D. H.\ 2004, ApJ, 609, 952
\bibitem[Zhang 2007]{2007ChJAA...7....1Z} Zhang, B.\ 2007, Chinese Journal of Astronomy and Astrophysics, 7, 1
\bibitem[Zhang et al. 2006]{2006ApJ...642..354Z} Zhang, B., Fan, Y.~Z., Dyks, J., Kobayashi, S., M{\'e}sz{\'a}ros, P., Burrows, D.~N., Nousek, J.~A., \& Gehrels, N.\ 2006, ApJ, 642, 354
\bibitem[Zhang et al.(2007)]{2007ApJ...666.1002Z} Zhang, B.-B., Liang, E.-W., \& Zhang, B.\ 2007, ApJ, 666, 1002
\bibitem[Zhang et al.(2008)]{2008arXiv0808.3793Z} Zhang, B.-B., Zhang, B., Liang, E.-W., \& Wang, X.-Y.\ 2008, arXiv:0808.3793
\bibitem{} Zhu, W. W., Xu, R. X., 2004, preprint (arXiv:astro-ph/0410265)
\end{thebibliography}
\end{document}